\documentclass{ifacconf}
 
\usepackage{graphicx}
\usepackage{natbib}
\usepackage{amsmath}
\usepackage{amssymb}
\usepackage{booktabs}
\usepackage{multirow}
\usepackage{array}
\usepackage{url}
\usepackage{tikz}
\usetikzlibrary{arrows.meta,positioning,shapes.geometric,calc,fit,backgrounds,decorations.pathreplacing}
\usepackage{xcolor}
 
\definecolor{plantBlue}{HTML}{1f4e79}
\definecolor{agentOrange}{HTML}{c75b12}
\definecolor{validatorGreen}{HTML}{2e7d32}
\definecolor{kgPurple}{HTML}{6a1b9a}
\definecolor{softGrey}{HTML}{f0f0f0}
\definecolor{midGrey}{HTML}{9e9e9e}
 
\begin{document}
 
\begin{frontmatter}
 
\title{A Tutorial on Autonomous Fault-Tolerant Control Using Knowledge-Grounded LLM Agents}
 
\thanks[footnoteinfo]{This research is funded by dtec.bw -- Digitalization
and Technology Research Center of the Bundeswehr (project ProMoDi).
dtec.bw is funded by the European Union -- NextGenerationEU. Financial
support from ABB for the Autonomous Industrial Systems Laboratory at
Imperial College London is gratefully acknowledged.\\
The first two authors contributed equally to this work.}
 
\author[First]{Javal Vyas}
\author[Second]{Milapji Singh Gill}
\author[Second]{Artan Markaj}
\author[Second]{Felix Gehlhoff}
\author[First]{Mehmet Mercang\"oz}
 
\address[First]{Autonomous Industrial Systems Laboratory,
                Imperial College London, London, United Kingdom
                (e-mail: j.vyas24@imperial.ac.uk).}
\address[Second]{Institute of Automation Technology,
                Helmut Schmidt University, Hamburg, Germany.}
 
\begin{abstract}
Fault recovery in process plants still relies heavily on plant
operators, especially when faults fall outside predefined supervisory
logic. Operators interpret alarms, procedures, P\&IDs, interlocks, and
process trends, then decide how to move the plant to a safe operating
mode without triggering a shutdown. This paper examines how Large
Language Model (LLM) agents can support such recovery decisions. The
proposed framework treats the LLM as a constrained supervisory
planner. It uses plant-specific knowledge to propose recovery actions,
and every proposal is checked by an external validator (symbolic or
simulation-based) before actuation. The paper develops three design
dimensions for applying the framework: the recovery patterns for which
LLM agents are useful, the validation strategies that separate
admissible from inadmissible proposals, and the deployment constraints
imposed by latency, knowledge engineering, safety integration, and
model lifecycle management. To make the framework directly usable, two
openly available executable Python environments are provided. Both
re-implement established case studies, a modular mixing module and a
continuous stirred-tank reactor, extended with configurable faults and
defined interfaces for custom recovery and validation methods.
\end{abstract}
 
 
\end{frontmatter}
 
\section{Introduction}
\label{sec:introduction}

Process plants are tightly coupled cyber-physical systems in which
faults such as actuator degradation, sensor bias, fouling, leaks, or
blocked flow paths can propagate quickly and move operation away from
its nominal regime. Active fault-tolerant control (FTC) aims to
preserve safe operation in the presence of such faults by detecting,
diagnosing, and counteracting their effects online
(\cite{Blanke.2016}). Classical FTC methods such as robust control,
supervisory reconfiguration, and model-based decision logic remain
indispensable, as they provide certified, deterministic responses to
the faults anticipated at design time. Robustness against
unanticipated faults, however, is more difficult to achieve, because
no pre-defined recovery is available. A system that can solve this problem must be capable of deciding autonomously at runtime, drawing on plant knowledge that is described
across heterogeneous artefacts such as P\&IDs, cause-effect matrices,
control logic, interlocks, operating envelopes, and informal
procedures (\cite{Webert.2022, RUPPRECHT2026101209}).

Large Language Models (LLMs) can reason over textual descriptions,
structured metadata, and operator-like explanations to produce
candidate recovery actions (\cite{RUPPRECHT2026101209,
BALDEA2025109064}). Their stochastic nature, however, makes them
unsuitable as direct controllers. They may hallucinate component
names, propose infeasible actions, or ignore actuator limits. The
question for the control community is therefore how to combine
classical FTC methods with LLMs, and how to test different
solutions. This requires a reusable framework with fixed interfaces
for knowledge retrieval, action generation, validation, and fallback
logic, so that different recovery and validation methods can be
compared under common assumptions.

This paper presents such a framework. The LLM acts as a
constrained supervisory planner whose proposals are grounded in
plant-specific knowledge and validated before actuation. A multi-agent
workflow decomposes operator duties into monitoring, planning, action
synthesis, simulation, validation, and reprompting, supported by a
plant knowledge graph that supplies relation-aware context and a
digital twin that evaluates proposed actions against process dynamics.
Low-level control remains with conventional controllers, so that the
LLM contributes semantic flexibility without holding execution
authority. Building on a concrete instantiation of this agent system presented in
our earlier work (\cite{GillVyas}), the present paper keeps the agents
and workflow fixed and instead addresses the design dimensions that
govern how the framework is applied. These dimensions are developed in
turn in Sections~\ref{sec:patterns}--\ref{sec:deployment}. To make the
framework directly usable, we provide two openly available, executable environments, a discrete mixing module and a continuous stirred-tank
reactor, that expose configurable faults, observable signals, and
interfaces for plugging in custom recovery and validation
logic\footnote{\url{https://github.com/AISL-at-Imperial-College-London/ctrl-alt-recover}}.
Rather than reporting performance figures, this paper focuses on what
must be considered and tested when adapting such a framework to one's
own application, while the repository documents the practical
deployment and extension of the environments through accompanying
tutorials.

The remainder of the paper is organized as follows.
Section~\ref{sec:related} reviews related work and positions the
contribution, and Section~\ref{sec:framework} recaps the framework and
introduces the three \emph{design dimensions} along which it is
applied. Sections~\ref{sec:patterns}--\ref{sec:deployment} develop
these in turn, Section~\ref{sec:environments} presents the two
executable environments for testing custom methods, and
Section~\ref{sec:conclusion} concludes.
 
\section{Related Work and Positioning}
\label{sec:related}

Active FTC for cyber-physical systems is a mature field that spans
observer-based residual generation, supervisory reconfiguration,
model predictive recovery, and learning-based control
(\cite{Blanke.2016, Olivier_etal_2017}). In parallel, a growing body
of work explores LLMs and agentic architectures for control-related
tasks at various levels of abstraction, including direct action
selection in building automation (\cite{HVACLLM}), supervisory
orchestration in industrial automation (\cite{Xia.11205539}), and operator-assistance interfaces (\cite{Sakhinana.2024}). Across this
body of work, each contribution is typically demonstrated on a single
task and a single plant. The language model is rarely integrated with
the safety mechanisms required for industrial fault recovery.
Proposals are seldom verified against plant dynamics or admissibility
constraints, and deployment aspects remain largely unaddressed. The
contribution of this paper is a framework and executable environments that can be useful for researchers working in this area.

We position our contribution along two axes. The first is an
\emph{application landscape}: active FTC problems differ in the layer
at which recovery acts (regulatory, supervisory, or planning), the
process dynamics (discrete-event, continuous, or hybrid), and the
degree to which fault semantics are enumerated at design time. An LLM
agent adds most where recovery requires interpreting plant knowledge
that is not exhaustively pre-enumerated, and where the action space is
supervisory rather than low-level. The application patterns of
Section~\ref{sec:patterns} make this concrete, including the cases in
which the framework is not the appropriate choice.

The second axis is the relation between LLM and plant. Following the
bounded-proposer principle (\cite{Vyas.2025, GillVyas}), the LLM is a
constrained supervisory planner without execution authority. Every
proposal is parsed into a structured schema and checked by an external
validator before actuation. The paper develops what this stance
demands in practice: where the framework applies, what to validate,
and how to deploy it, supported by the executable environments
introduced later.
 
\section{Framework and Design Dimensions}
\label{sec:framework}

\subsection{Framework}

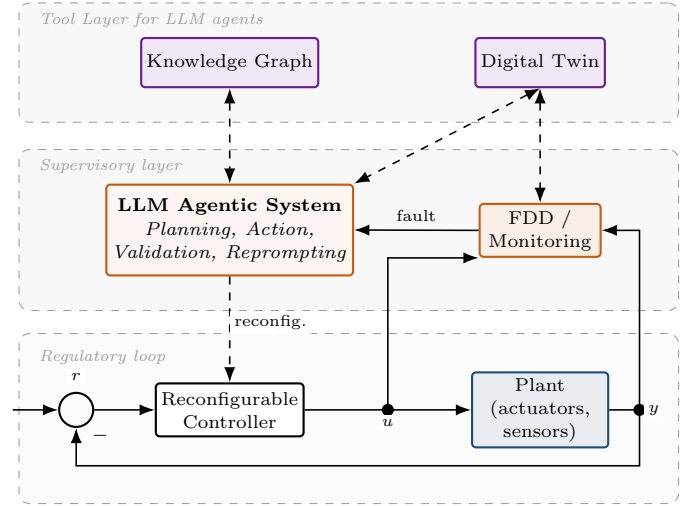
\begin{figure}[t]
\centering
\begin{tikzpicture}[
    font=\scriptsize,
    >=Latex,
    block/.style={
        rectangle, draw, thick, rounded corners=1.5pt,
        minimum height=7mm, align=center, fill=white, inner sep=2pt
    },
    ctrl/.style={block, minimum width=18mm},
    plant/.style={block, draw=plantBlue, fill=plantBlue!10, minimum width=18mm},
    fdd/.style={block, draw=agentOrange, fill=agentOrange!10, minimum width=16mm},
    sup/.style={block, draw=agentOrange, fill=agentOrange!6},
    tool/.style={block, draw=kgPurple, fill=kgPurple!10,
                 minimum width=16mm, minimum height=6.5mm},
    sumc/.style={circle, draw, thick, inner sep=0pt, minimum size=4.5mm},
    sig/.style={->, semithick},
    line/.style={semithick},
    rec/.style={->, semithick, dashed},
    sigLab/.style={font=\tiny, inner sep=1.5pt, fill=white},
    branch/.style={circle, fill=black, inner sep=0pt, minimum size=1.6mm},
    grp/.style={draw=midGrey, dashed, rounded corners}
]

\node[sumc] (sum) at (0,0) {};
\node[ctrl, right=8mm of sum] (controller) {Reconfigurable\\Controller};
\node[plant, right=22mm of controller] (plant) {Plant\\(actuators,\\sensors)};

\draw[sig] ($(sum.west)-(6mm,0)$) -- (sum.west);
\node[sigLab, above=1mm of sum] {$r$};

\draw[sig] (sum) -- (controller);

\coordinate (ubranch) at ($(controller.east)!0.5!(plant.west)$);
\draw[line] (controller.east) -- (ubranch);
\draw[sig]  (ubranch) -- (plant.west);
\node[branch] at (ubranch) {};
\node[sigLab, below=0.6mm of ubranch] {$u$};

\coordinate (ybranch) at ($(plant.east)+(4mm,0)$);
\draw[line] (plant.east) -- (ybranch);
\node[branch] at (ybranch) {};
\node[sigLab, right=0.6mm of ybranch] {$y$};
\draw[sig] (ybranch) |- ($(sum.south)+(0,-5mm)$) -| (sum.south);
\node[font=\scriptsize\bfseries, inner sep=0pt, anchor=north west]
    at ($(sum.south)+(2.0mm,-0.0mm)$) {$-$};

\node[fdd, above=15mm of plant] (fdd) {FDD /\\Monitoring};
\draw[sig] (ubranch) |- (fdd.south west);
\draw[sig] (ybranch) |- (fdd.east);

\node[sup, above=14mm of controller, minimum width=30mm, minimum height=12mm]
    (llm) {\textbf{LLM Agentic System}\\\textit{Planning, Action,}\\\textit{Validation, Reprompting}};

\draw[sig] (fdd.west) -- (llm.east);
\node[font=\tiny, inner sep=1pt] at ($(fdd.west)!0.5!(llm.east)+(0,2mm)$) {fault};

\draw[rec] (llm.south) -- (controller.north)
    node[pos=0.43, right, sigLab] {reconfig.};

\node[tool] (kg)   at ($(llm.north)+(0,16mm)$) {Knowledge Graph};
\node[tool] (twin) at (kg -| fdd)              {Digital Twin};

\draw[rec,<->] (kg.south)   -- (llm.north);
\draw[rec,<->] (twin.south) -- (fdd.north);

\draw[rec,<->] (twin.south) -- (llm.north east);

\begin{pgfonlayer}{background}
    \coordinate (Lx) at ($(sum.west)+(-4mm,0)$);
    \coordinate (Rx) at ($(ybranch)+(4mm,0)$);

    \node[grp, fill=softGrey!25,
        fit={(Lx |- controller.south) (Rx |- controller.south)
             (controller.north) (plant) (sum)
             ($(sum.south)+(0,-5mm)$)},
        inner ysep=5mm] (reglayer) {};

    \node[grp, fill=softGrey!45,
        fit={(Lx |- llm.south) (Rx |- llm.south) (llm.north) (fdd)},
        inner ysep=4.5mm] (suplayer) {};

    \node[grp, fill=softGrey!45,
        fit={(Lx |- kg.south) (Rx |- kg.south) (kg.north) (twin)},
        inner ysep=4.5mm] (toollayer) {};
\end{pgfonlayer}

\node[font=\tiny\itshape, midGrey, anchor=north west]
    at ($(toollayer.north west)+(1.5mm,-0.1mm)$) {Tool Layer for LLM agents};
\node[font=\tiny\itshape, midGrey, anchor=north west]
    at ($(suplayer.north west)+(1.5mm,-0.1mm)$) {Supervisory layer};
\node[font=\tiny\itshape, midGrey, anchor=north west]
    at ($(reglayer.north west)+(1.5mm,-1mm)$) {Regulatory loop};

\end{tikzpicture}
\caption{The proposed framework cast onto the canonical active
fault-tolerant control loop.}
\label{fig:overview}
\end{figure}

This subsection recaps the agentic framework that underlies the
present paper. It condenses the authors' prior work
(\cite{GillVyas, Vyas.2025}), to which the reader is referred for the
concrete realisation of the agents, their interaction, and the
underlying knowledge representation. Here we describe it only at an
abstract level needed for the application-oriented discussion that
follows, using the canonical active FTC loop of
Figure~\ref{fig:overview}.

The framework casts an LLM agentic system onto this loop and
distinguishes a physical and a virtual space. The regulatory loop,
consisting of the reconfigurable controller and the plant, is left
untouched and continues to operate under conventional control. The LLM agentic system occupies the supervisory reconfiguration
level: it consumes the fault context produced by the fault detection
and diagnosis (FDD) component and adapts the controller, shown as a
dashed reconfiguration link in Figure~\ref{fig:overview}. 
The agents decompose operator duties into
monitoring, planning, action synthesis, simulation, validation, and
reprompting, and consult a tool layer: a knowledge graph that grounds
them in plant-specific context, and a digital twin that evaluates proposed
actions against process dynamics. The same tool layer can also support
fault detection and monitoring, for instance through model-based
residuals or structural diagnosis knowledge. Throughout, solid arrows
denote signal flow and dashed arrows denote reconfiguration and
supervisory adaptation.

This loop can be made precise as follows. Let $x_t$ denote the plant
state at supervisory time $t$, $f_t$ the detected fault context,
$\mathcal{K}$ the knowledge graph, and $\mathcal{A}$ the admissible
supervisory action space. The agents construct a prompt
\begin{equation}
    p_t = \Phi\big(x_t,\, f_t,\, R(\mathcal{K}, x_t, f_t),\, h_t\big),
    \label{eq:prompt}
\end{equation}
where $R(\cdot)$ retrieves relevant context from $\mathcal{K}$ and
$h_t$ holds validation feedback from previous attempts. The LLM then
proposes a candidate action,
\begin{equation}
    \hat{a}_t \sim \pi_\theta(\cdot \mid p_t), \qquad
    \hat{a}_t \in \mathcal{A},
    \label{eq:propose}
\end{equation}
which an external validator judges before any actuation,
\begin{equation}
    z_t = V(x_t,\, \hat{a}_t,\, \mathcal{K},\, \mathcal{M}),
    \label{eq:validate}
\end{equation}
where $\mathcal{M}$ is either a deterministic transition model or a
digital twin. The action is applied only if $z_t = \mathrm{accept}$.
Otherwise the failure reason is appended to $h_t$ and the LLM is
reprompted via \eqref{eq:prompt} until a fixed budget is exhausted, at
which point control passes to an engineered safety fallback.

While the agents and the overall workflow are fixed, the framework
deliberately leaves open how its remaining elements are realised. The
knowledge graph can be built on a standards-based ontology
(\cite{Gill.9920259122025}), and the retrieval $R(\cdot)$ may be
implemented through graph-based retrieval-augmented generation or
other means. Likewise, the validator $V$ and the model $\mathcal{M}$
are treated as components that the user instantiates for the
application at hand.

\subsection{Design dimensions}

Concretising the framework for a given application means making
choices along three \emph{design dimensions}, each tied to a property
of the plant and each motivating one of the following sections.

\textbf{Dimension 1 -- The recovery problem shapes the agent's role.}
Whether recovery means selecting a discrete operating mode, adjusting
continuous setpoints, or both, determines the action space
$\mathcal{A}$ the agent operates on and the errors it can make. This
is the basis for the \emph{application patterns} of
Section~\ref{sec:patterns}.

\textbf{Dimension 2 -- The plant defines what is admissible.}
Which transitions, envelopes, interlocks, and transient responses are
acceptable is a plant-specific question that the framework delegates
to the external validator $V$, whose type must match the recovery
problem rather than being a free choice. This is the subject of the
\emph{validation strategies} of Section~\ref{sec:validation}.

\textbf{Dimension 3 -- The plant sets the resource budget.}
The knowledge graph must be built for a specific plant, and the
agentic loop must decide within the latency the process allows: the
delivered value is bounded by the completeness of the knowledge, the
admissible compute placement by the process dynamics. These are the
\emph{deployment considerations} of Section~\ref{sec:deployment}.
 
\section{Application Patterns}
\label{sec:patterns}

As introduced in the first design dimension
(Section~\ref{sec:framework}), the applicability of the framework
depends less on the underlying control theory than on how the
recovery problem is structured. Process plants combine two regimes:
discrete supervisory logic governing operating modes, sequences, and
routing, and continuous regulatory control holding variables at their
setpoints. Faults may require intervention in either regime, and this
distinction, rather than the specific unit operation, determines what
an LLM agent must do and how its proposals must be checked. It yields
three recurring patterns, each characterized by its action space, the
validator that fits naturally, and the dominant failure modes shown in
Table~\ref{tab:patterns}.

\textbf{Pattern A -- Discrete supervisory routing.}
The plant evolves through a finite set of operating modes captured as
a state machine, as in batch sequences, modular skids, or the
supervisory layer of a continuous plant. A fault changes which
transitions remain admissible, so recovery means choosing a
fault-consistent route to a safe state and activating the actuators
each visited state requires. The action space is combinatorial but
small, and admissibility is a property of the state-machine graph. An
LLM agent suits this pattern because route selection benefits from
natural-language fault semantics that are hard to enumerate
exhaustively in advance.

\textbf{Pattern B -- Continuous setpoint adaptation.}
The plant runs under regulatory control, typically PID, and recovery
adjusts a small set of supervisory setpoints rather than reconfiguring
the low-level loops. The action space is continuous but searched over bounded ranges and discrete increments. Admissibility is a matter of
dynamics, not syntax: a setpoint can be numerically valid yet
infeasible under closed-loop behaviour, for instance when a
temperature target is unreachable because cooling authority is
saturated. An LLM agent helps because deciding the trade-off to
enforce, such as reducing throughput rather than demanding more
cooling, is a high-level judgement over plant knowledge.

\textbf{Pattern C -- Hybrid recovery.}
Most real plants combine both: a fault first triggers a supervisory
mode change (Pattern~A), which then requires re-tuning setpoints
within the new mode (Pattern~B). Recovery is staged, with a symbolic
check for the mode change and a simulation-based one for the
setpoints. The consequence is architectural: the agentic loop must
track which stage it is in, so that feedback from one stage is not
misread as a failure of the other.

\begin{table}[t]
\caption{Application patterns and their validation regimes.}
\label{tab:patterns}
\centering
\small
\begin{tabular}{@{}l p{0.20\columnwidth} p{0.25\columnwidth} p{0.23\columnwidth}@{}}
\toprule
\textbf{Pattern} & \textbf{Action space} & \textbf{Validator (cost)} &
\textbf{Failure modes} \\
\midrule
A & Finite states + actuator sets & Symbolic; near-free, many
reprompts & Wrong route, actuator mismatch, hallucinated entity \\
B & Bounded setpoints, discrete increments & Simulation-based rollout;
costly, small budget & Dynamic infeasibility, actuator saturation,
slow recovery \\
C & Staged: mode + setpoints & Both, applied in stages & Stage
confusion, premature feedback, mis-attribution \\
\bottomrule
\end{tabular}
\end{table}

The patterns are not exhaustive, but they cover a large fraction of
supervisory-recovery problems, and the two environments of
Section~\ref{sec:environments} instantiate Patterns~A and~B directly.
They also make explicit when the framework is \emph{not} the right
tool: when a rule base already captures the problem and Pattern~A
degenerates to a lookup, when the dynamics demand millisecond-level
reaction so that the LLM call dominates the time budget, or when the
fault space is narrow enough that deterministic reconfiguration
suffices.
 
\section{Validation Strategies}
\label{sec:validation}

As set out in the second design dimension
(Section~\ref{sec:framework}), the plant, not the LLM, defines what
counts as an admissible recovery, and the validator is where this
judgement is enforced. It is the boundary that turns a stochastic LLM
proposal into an engineered control action, and thus the single most
consequential design decision in the framework. It determines what classes of invalid proposals are detected, what each recovery attempt costs, and how the agentic loop is structured. This section compares validation strategies by detection scope, timing, and cost.

\subsection{Symbolic versus simulation-based validation}

A \emph{symbolic validator} checks a proposal against the structure of the plant: whether the next state is reachable, whether
the named actuators exist, and whether the actuator pattern is
consistent with the selected state. Such checks are inexpensive,
deterministic, and catch all admissibility violations encoded in the symbolic model, but they do not see dynamics. A syntactically admissible
proposal may still be dynamically infeasible, for instance a
temperature setpoint the cooling actuator cannot reach once saturated.

A \emph{simulation-based validator} clones the plant state into a
digital twin, applies the action, and rolls the dynamics forward,
accepting only if the rollout meets criteria such as bounded unsafe
exposure, time to a safe state, persistence of recovery, and respect
for actuator saturation. It catches what symbolic checks miss, but at
a higher cost, since even a short rollout dominates the per-iteration
latency. Which validator fits which pattern follows directly from this
distinction and is summarised in Table~\ref{tab:patterns}.

Validation also has a temporal axis. Checking a proposal before issuing any actuator command is the conservative choice and fits supervisory intervention, where recovery actions are discrete and infrequent. When the action unfolds over a longer window, such as a slow setpoint ramp, runtime monitoring is added to detect deviations between the predicted and the actual trajectory. It shares the digital-twin infrastructure but runs on a different trigger, since the rollout is computed once before execution while monitoring evaluates the live trajectory continuously.

\subsection{Cost--latency profile and the reprompt loop}

The validator type also determines the cost-latency profile of the
loop and the affordable reprompt budget. Symbolic validation is
essentially free, so many cycles are affordable. Simulation-based
validation is expensive, so the budget must be small. This places a
premium on feedback quality. A structured response converges much
faster than a generic rejection. It names which envelope was breached,
by how much, and on which actuator. Bounding reprompts by repeated
failure category, rather than by raw count, keeps cost down without
weakening the safety contribution.
 
\section{Deployment Considerations}
\label{sec:deployment}

A framework that is sound in the laboratory still faces hurdles before
it is useful in a process plant, and these lie less in the LLM itself
than around it. We group them into four concerns, each a potential
showstopper in its own right and each examinable in the executable
environments of Section~\ref{sec:environments}.

\subsection{Compute and latency}

Cloud-hosted LLMs offer the strongest models, but add network
latency, expose process information to third parties, and tie
recovery to an external service-level agreement. On-premise or edge
models eliminate these issues but offer lower reasoning capability
for the same cost. The choice is governed by the latency budget,
which the process dynamics set rather than the LLM. Batch and slow
supervisory tasks, with per-decision latency budgets of tens of
seconds, tolerate cloud LLMs comfortably. Tightly regulated
continuous loops, by contrast, require on-premise serving or
caching. As in Pattern~B, the loop can also be split into a slow
supervisory part where the LLM operates and a fast regulatory part
that remains under conventional control.

\subsection{Knowledge engineering effort}

The knowledge graph that grounds the LLM is, in practice, the dominant
deployment cost. For a new plant it requires collecting and
normalising engineering artefacts such as P\&IDs, interlock lists,
operating envelopes, and fault catalogues, then aligning them with an
ontology such as the standards-based one in (\cite{Gill.9920259122025}).
Two properties ease this. The schema is plant-agnostic and reusable
across assets, so only the instance data is populated per plant. Moreover,
although automated extraction from engineering documents is an active
research direction, current methods still require human-in-the-loop
review for safety-critical artefacts. Knowledge engineering is
therefore best treated as a one-time integration cost per plant, with
the schema designed for reuse rather than for any particular fault.

\subsection{Safety integration}

An LLM agent must coexist with existing safety layers such as
interlocks, safety instrumented systems, and operator procedures, and
three principles keep it subordinate to them. First, the LLM never
holds execution authority. All proposals pass through the validator
and, if rejected, through the engineered fallback, preserving the
existing defence-in-depth. Second, every proposal is logged as a
structured action together with its retrieved context, the validator
verdict, and the 
outcome. This decision trace is the basis
for audit and post-incident review and is, in our view, non-negotiable
for industrial use. Third, the IT/OT boundary must be explicit, since
prompt injection through manipulated inputs becomes a credible threat
once the LLM consumes plant text.

\subsection{Model lifecycle}

Finally, the LLM itself is a moving target. Cloud models are versioned
by the provider but may still change without notice, while on-premise
models change when the operator updates them. Three practices help
keep behaviour reproducible. First, the model identifier is pinned and
logged with every decision. Second, a regression suite of recorded
fault scenarios is replayed on every update, with acceptance gated on
the validator verdict rather than on textual similarity. Third, the
structured output schema is treated as a stable contract, because
schema breakage is more likely than reasoning regression and easier to
detect automatically.

In practice, these concerns reduce to a short readiness check before
an instance is moved to a higher-fidelity environment. The knowledge
graph should be complete for the fault scenarios of interest, with
human-reviewed mappings and interlocks. The validator should be
matched to the application pattern, with a documented cost-latency
budget. Every executed action should have an auditable decision trace
and a documented fallback. The LLM identifier should be pinned, with a
regression suite and an explicit IT/OT boundary.

\section{Executable Case-Study Environments}
\label{sec:environments}

Two openly available Python environments accompany the framework as a
testbed for the design dimensions of
Sections~\ref{sec:patterns}--\ref{sec:deployment}. They build on a
modular cyber-physical production system (\cite{J.Ehrhardt.2022}) and
a stirred-tank reactor training environment (\cite{Markaj.2024}),
re-implemented with fault injection and defined interfaces for
external recovery and validation logic. We describe each in nominal
operation, with its injectable faults and plug-in points, rather than
reporting performance results. Together they span the two ends of the
pattern spectrum: discrete supervisory routing and continuous setpoint
adaptation.

\subsection{A discrete environment: the mixing module}
\label{sec:env-mixer}

\begin{figure}[t]
    \centering
    \includegraphics[width=\linewidth]{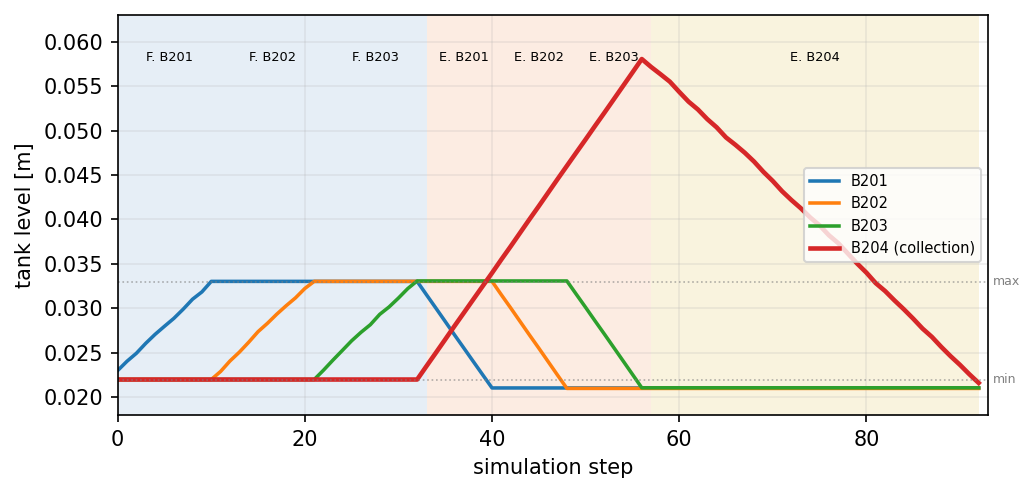}
    \caption{Mixing module under nominal operation: the three batch
    tanks (B201--B203) are filled in turn and then emptied into the
    collection tank (B204), which is drained last. Shaded bands mark
    the active state.}
    \label{fig:mixer-behaviour}
\end{figure}

The first environment is a modular mixing unit with three batch tanks
(B201--B203) and a collection tank (B204), realised in Python as a
finite state machine. Nominal operation, shown in
Figure~\ref{fig:mixer-behaviour}, fills the three tanks in turn and
then empties them one after another into the collection tank, which is
finally drained. The controllable elements are on--off valves and two
pumps, a main pump and a bypass pump, so the action space is discrete
and small. This makes the environment an instance of Pattern~A:
recovery means selecting a fault-consistent route through the state
machine and activating the actuators each visited state requires.

The injectable faults change which route remains admissible. A pump
failure, pump degradation, a clogged main line, or a leak render the
main-pump emptying path unusable, so the agent must switch to the
bypass-pump route. A sensor fault, by contrast, leaves the main route
valid, and switching to the bypass would be an unnecessary
over-reaction. Distinguishing faults by the route they invalidate
rather than by numerical severity keeps the recovery problem symbolic.
The observable signals are the tank levels, the current state, and
the active actuator configuration. This environment is the natural
place to exercise a symbolic validator
(Section~\ref{sec:validation}), since the state-machine structure and
the actuator mapping are directly available. A custom planner plugs
in at a defined interface. It consumes the typed fault and returns a
target state with its actuator set.

\subsection{A continuous environment: the stirred-tank reactor}
\label{sec:env-cstr}

\begin{figure}[t]
    \centering
    \includegraphics[width=\linewidth]{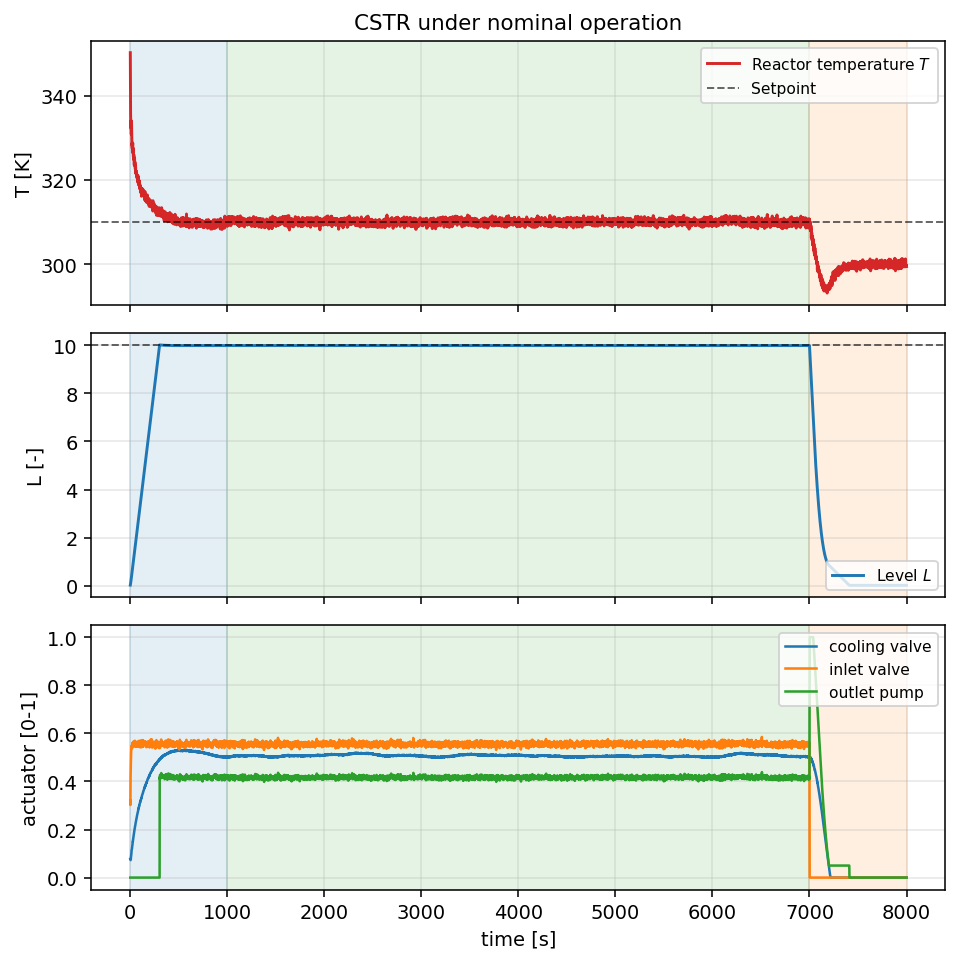}
    \caption{CSTR under nominal operation: temperature and level are held
at their setpoints across the start-up, normal, and shut-down phases
(shaded), with actuators well below saturation.}
    \label{fig:cstr-behaviour}
\end{figure}

The second environment is a continuous stirred-tank reactor (CSTR)
under regulatory PID control. It is re-implemented in Python from the
training environment of \cite{Markaj.2024} and simulates the coupled
mass and energy balances of an exothermic reaction. The plant passes
through start-up, normal, and shut-down phases. Under nominal
operation, shown in Figure~\ref{fig:cstr-behaviour}, three PID loops
hold the reactor temperature near its setpoint, the level constant,
and the inlet flow steady. The cooling, inlet, and outlet actuators
settle at moderate positions with ample margin. Here the controllable
elements are not discrete commands but a small set of supervisory
setpoints: reactor temperature, level, and inlet flow. The environment
is therefore an instance of Pattern~B. Recovery means proposing
setpoint adaptations that keep the process safe despite the fault.

The injectable faults are chosen to expose the dynamic character of
continuous recovery. Fouling gradually reduces the heat-transfer
capability, so the cooling valve opens further over time and
eventually saturates, after which the temperature can no longer be
held by cooling alone and recovery requires reducing the thermal load
by lowering the inlet-flow setpoint. Pump degradation reduces
outlet-flow authority and causes the level to drift, while a cooling
valve constrained near its closed position caps cooling authority in
the same way as fouling. The observable signals are the continuous
temperature and level measurements, the inlet, outlet, and cooling
flows, the actuator positions, and derived quantities such as the
mass-balance residual. This environment is the natural place to
exercise a simulation-based validator
(Section~\ref{sec:validation}), since a proposed setpoint can be
numerically valid yet dynamically infeasible. The environment itself
serves as the digital twin against which a proposal is rolled out
before acceptance. A custom recovery method is plugged in at the
interface that consumes the current process snapshot and the fault
context and returns a setpoint triple.

\section{Conclusion}
\label{sec:conclusion}

In this contribution, a knowledge-grounded LLM agentic framework for
active FTC was presented from an
application-oriented perspective, organized along three design
dimensions: the application patterns it fits, the validation
strategies that decide admissibility and cost, and the deployment
considerations that constrain practical use. Two openly available
executable environments are provided, in which custom recovery and
validation methods can be evaluated. Each run records the fault
context, the proposed action, the validator verdict, and the resulting
trajectory. The three dimensions can therefore be studied directly,
and a method transfers to a higher-fidelity plant model through the
same interfaces. Across all three, the validator, the knowledge graph,
and the integration with existing safety layers shape practical
behaviour more than the choice of model family. Future work will
extend the patterns to multi-unit recovery, study automated
knowledge-graph construction with human-in-the-loop verification, and
examine how validator feedback shapes the reprompt loop across broader
fault classes.
 
\bibliography{cas-refs}

\end{document}